\documentclass[sigconf, 9pt]{acmart}

    \PassOptionsToPackage{numbers, compress}{natbib}

\usepackage[utf8]{inputenc} 
\usepackage[T1]{fontenc}    
\usepackage{hyperref}       
\usepackage{url}            
\usepackage{booktabs}       
\usepackage{amsfonts}       
\usepackage{nicefrac}       
\usepackage{microtype}      
\usepackage{graphicx, color}
\usepackage{subfigure}
\usepackage{amsmath}
\usepackage{amssymb}
\usepackage{amsthm}
\usepackage{mathtools}

\usepackage{color}

\AtBeginDocument{
  \providecommand\BibTeX{{%
    \normalfont B\kern-0.5em{\scshape i\kern-0.25em b}\kern-0.8em\TeX}}}

\acmYear{2019}
\copyrightyear{2019}
\setcopyright{rightsretained}
\acmConference[CoNEXT '19 Companion]{The 15th International Conference on
emerging Networking EXperiments and Technologies}{December
9--12,2019}{Orlando, FL, USA} 
\acmBooktitle{The 15th International
Conference on emerging Networking EXperiments and Technologies (CoNEXT '19
Companion), December 9--12, 2019, Orlando, FL, USA} 
\acmPrice{15.00}
\acmDOI{10.1145/3360468.3368176} 
\acmISBN{978-1-4503-7006-6/19/12}

\begin{document}
\title{One Pixel Image and RF Signal Based Split Learning for mmWave Received Power Prediction}
\author{Yusuke Koda}
\email{koda@imc.cce.i.kyoto-u.ac.jp}
\affiliation{%
  \institution{Graduate School of Informatics, Kyoto University, Kyoto, Japan}
}
\author{Jihong~Park}
\email{jihong.park@oulu.fi}
\affiliation{%
  \institution{Centre for Wireless Communication, University of Oulu, Oulu, Finland}
}
\author{Mehdi~Bennis}
\email{mehdi.bennis@oulu.fi}
\affiliation{%
\institution{Centre for Wireless Communication, University of Oulu, Oulu, Finland}}

\author{Koji Yamamoto}
\email{kyamamot@i.kyoto-u.ac.jp}
\affiliation{%
  \institution{Graduate School of Informatics, Kyoto University, Kyoto, Japan}
}
\author{Takayuki Nishio}
\email{nishio@i.kyoto-u.ac.jp}
\affiliation{%
  \institution{Graduate School of Informatics, Kyoto University, Kyoto, Japan}
}
\author{Masahiro Morikura}
\email{morikura@i.kyoto-u.ac.jp}
\affiliation{%
  \institution{Graduate School of Informatics, Kyoto University, Kyoto, Japan}
}

\begin{abstract}
 Focusing on the received power prediction of millimeter-wave (mmWave) radio-frequency (RF) signals, we propose a \emph{multimodal split learning (SL)} framework that integrates RF received  signal powers and depth-images observed by physically separated entities. To improve its communication efficiency while preserving data privacy, we propose an SL neural network architecture that compresses the communication payload, i.e., images. Compared to a baseline solely utilizing RF signals, numerical results show that SL integrating only \emph{one pixel image} with RF signals achieves higher prediction accuracy while maximizing both communication efficiency and privacy guarantees.
\end{abstract}

\maketitle
\section{Introduction}

Received power prediction of radio frequency (RF) signals in millimeter wave (mmWave) is a key enabler for proactive 5G cellular system operations, and is a daunting task~\cite{MehdiURLLC:18}. 
These signals are sensitive to physical blockages, and the received power levels drop significantly once the signals are blocked, i.e., in non-line-of-sight (non-LoS) channel conditions \cite{JHParkTWC:15}. 
When either LoS or non-LoS conditions are consistent,
 the sudden variation of power levels give almost no prior indications in the RF signal domain, which motivates us to utilize non-RF domains that include useful features.

Depth-camera images contain suitable features for this purpose. 
Observing a sequence of depth-image frames 
enables detecting signs of transitions between LoS and non-LoS conditions in millimeter wave (mmWave) links \cite{nishio_jsac}. 
However, their captured images should be exchanged over wireless links, which may incur huge communication overhead while violating data privacy. 

Spurred by this problem, we propose a communication-efficient and privacy-preserving \emph{multimodal split learning (SL)} framework for mmWave power prediction (see Fig.~1), in which a global model is split into two parts: convolutional neural network (CNN) layers processing depth-camera images and recurrent neural network (RNN) layers combining the CNN outputs with RF signal inputs, which are connected over wireless links. The key idea is to compress the CNN output dimension by adjusting the pooling region size, thereby achieving lower inter-layer communication overhead while preserving more data privacy.


\begin{figure}[t]
  \centering
  \includegraphics[width=0.5\columnwidth]{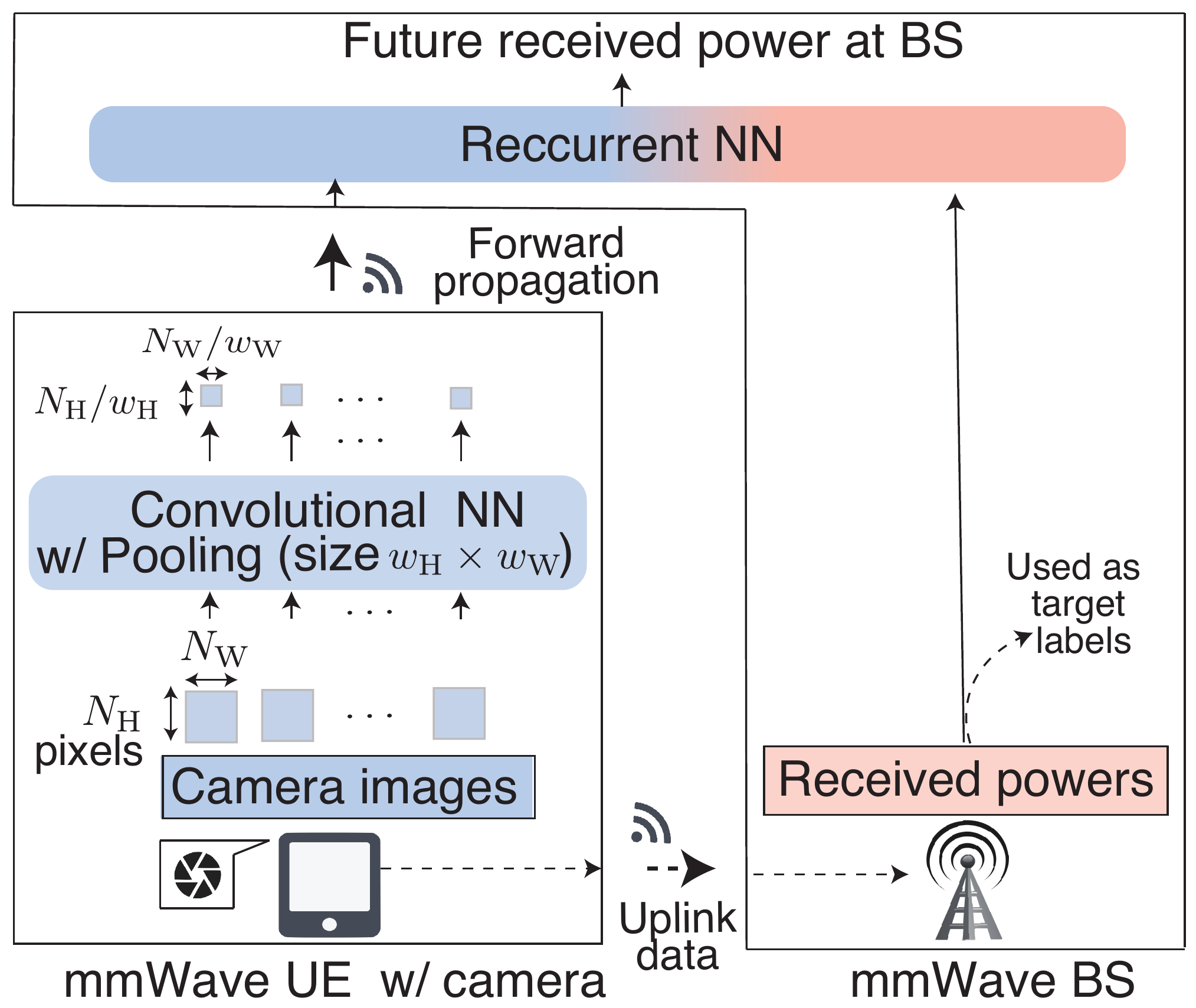}
  \vspace{-5pt}
  \caption{\small Multimodal SL architecture that integrates image and RF signal (Img+RF) features for predicting mmWave received power.
  }
  \label{fig:system_model}
\end{figure}
\begin{figure}[t]
  \centering
  \subfigure[Raw images.]{\includegraphics[width=0.1\textwidth]{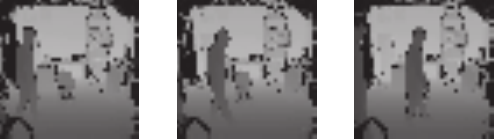}}\hspace{10pt}
    \subfigure[$w_\text{H}\!\times\! w_\text{W}$: $1\times 1$.]{\includegraphics[width=0.1\textwidth]{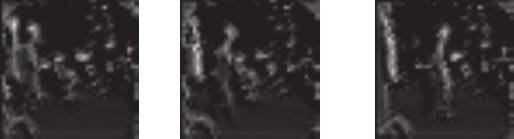}}\hspace{10pt}
    \subfigure[$4\!\times\! 4$.]{\includegraphics[width=0.1\textwidth]{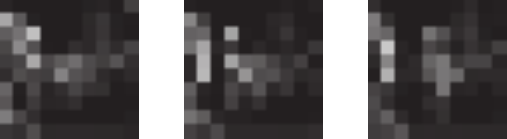}}\hspace{10pt}
    \subfigure[$40\!\times\! 40$ (\textbf{1-pixel}).]{\includegraphics[width=0.1\textwidth]{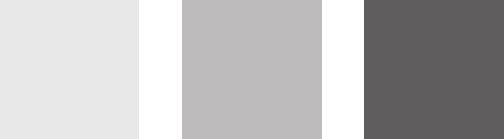}} \vspace{-2em}
    \caption{Raw depth-images and CNN output images.} \label{fig:images}
  \end{figure}

\section{System Model}
\textit{\textbf{Multi-modal SL architecture.}} Our proposed multimodal SL neural network architecture is summarized in Fig.~\ref{fig:system_model}.
We consider an mmWave user equipment (UE) transmitting mmWave signals to a base station (BS), while observing this uplink channel using a depth-camera. 
The BS predicts this uplink received signal power with the following SL architecture. 
The UE stores CNN layers processing raw depth-images with width $N_W$ and height $N_H$, and the CNN output is feed-forwarded to the BS through a wireless link. The BS runs RNN layers whose output is the future received power prediction, and the input is a concatenated sequence of received RF signal powers and CNN output images. To compress the communication payload, the CNN output is filtered by an average pooling layer with dimension $w_{\mathrm{W}}\times w_{\mathrm{H}}$, resulting in the output feature map whose dimension is $(N_{\mathrm{W}}/w_{\mathrm{W}})\times (N_{\mathrm{H}}/w_{\mathrm{H}})$,
 i.e., $w_{\mathrm{W}}\times w_{\mathrm{H}}$ times compressed images (see Fig.~\ref{fig:images}).

\textit{\textbf{Wireless Channel Model.}} 
Feeding forward and backward propagation in the proposed SL architecture incur wireless uplink and downlink communications, respectively. Let the superscript $x\in\{\mathrm{UL}, \mathrm{DL}\}$ identify uplink and downlink. For the uplink and downlink, transmitters utilize frequency bandwidth $W^{(x)}$ and transmit power $P^{(x)}$. For a given distance $r$ between the BS and the UE, the received signal-to-noise ratio (SNR) at the $t$-th time slot is given by: $\mathit{SNR}^{(x)}_{t} = P^{(x)}r^{-\alpha} h^{(x)}_{t}/(\sigma^2 W^{(x)})$, where $\sigma^2$ denotes the noise power spectral density, and $\alpha$ is the path loss exponent. The term $h^{(x)}_{t}$ describes multi-path channel fading, which
 is an exponential random variable with unitary mean, independent and identically distributed across time slots. 
 Each received signal is successfully decoded if $\mathit{SNR}^{(x)}_{t} > 1 - 2^{B^{(x)}/(\tau W^{(x)})}$ 
  at the $t$-th time slot for a payload size $B^{(x)}$ and a unit time slot length~$\tau$; otherwise, the signals are re-transmitted in the next time slots as done in~\cite{ParkWCL:18}. 
 The uplink payload size is determined by the CNN output size as: $B^{(\mathrm{UL})} = N_{\mathrm{H}}N_{\mathrm{W}}BRL/(w_{\mathrm{H}}w_{\mathrm{W}})$, where $B$, $R$, and $L$ denote the mini-batch size, bit-depth, and the length for images, respectively.

\begin{figure}
  \centering
  \subfigure[Learning curves.]{\centering\includegraphics[width=0.44\columnwidth]{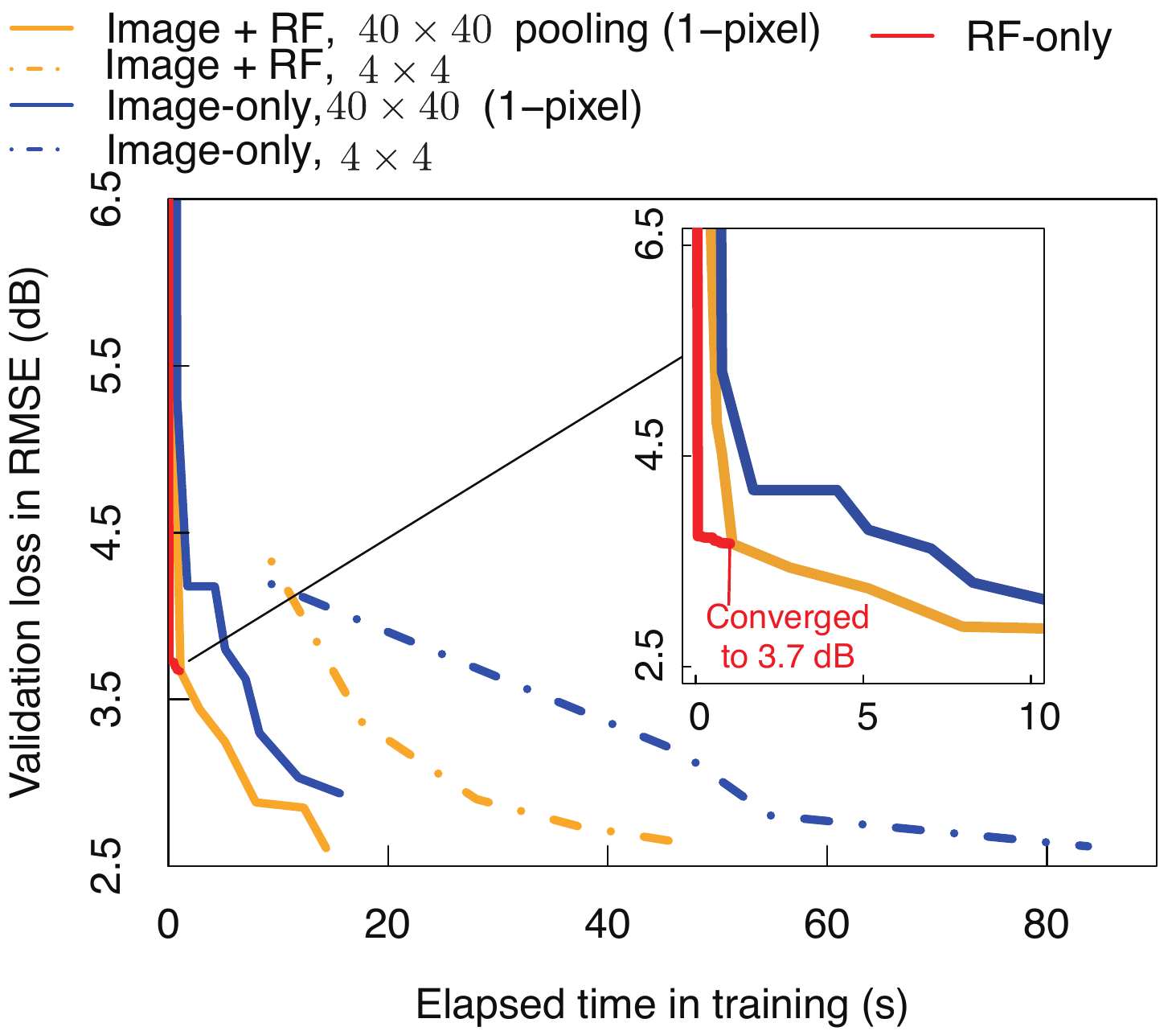}}
  \subfigure[Received power predictions.]{\centering\includegraphics[width=0.44\columnwidth]{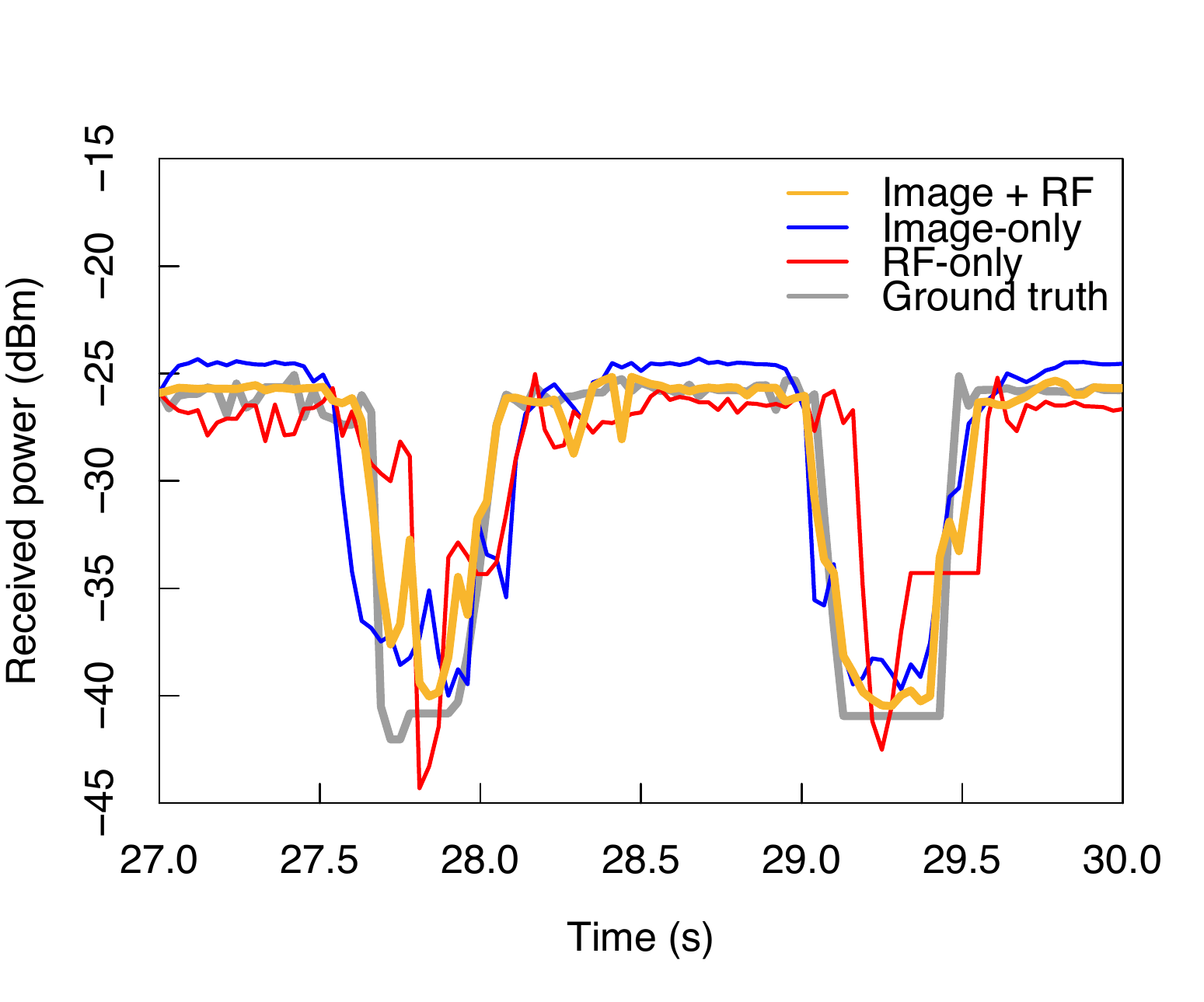}}
  \caption{Accuracy comparison via (a) learning curves and (b) received power predictions compared to the ground truth.}
  \label{fig:results}\vskip -15pt
\end{figure}

\section{Experiments}
\label{sec:ex}
\textit{\textbf{Dataset and Times-Series Prediction.}} Our dataset of depth-images and received mmWave power samples follows from \cite{koda_measurement2,nishio_jsac}, which is constructed by an experiment using a 60.48 GHz mmWave transmitter and a Microsoft Kinect depth camera. Each sample is denoted as $s_k = (x_k, P_k)_{k \in\mathcal{K}}$ where $x_k$ is the output image of CNN and $P_k$ is the corresponding received power. The index $k\in\mathcal{K}$ identifies the time index, where $\mathcal{K} = \{1, 2, \dots, 13,\!228\}$.
At the $k$-th time index, the RNN input layer of our proposed multimodal SL is fed by a sequence $\{s_{k-L+1},\dots,s_{k-1},s_k\}$ with length $L=4$, so as to predict the $T=120$ ms ahead received power $P_{k + T/\gamma}$, where $\gamma=33$ ms is given by the depth camera frame rate.

\textit{\textbf{Training and Validation.}} Let the training and validation datasets be denoted by $\mathcal{K}_{\mathrm{train}}=\{L, L + 1, \dots, 9,\!928\}$ and $\mathcal{K}_\mathrm{val}=\mathcal{K}\setminus\mathcal{K}_{\mathrm{train}}$, respectively. 
The loss function is the mean squared error (MSE) given as: $\sum_{k\in \mathcal{B}}(\hat{P}_{k + T/\gamma} - P_{k + T/\gamma})^2/|\mathcal{B}|$, where $\hat{P}_{k + T/\gamma}$ represents the predicted received power, and $\mathcal{B}$ is a minibatch uniformly randomly sampled from $\mathcal{K}_{\mathrm{train}}$. 
The loss is minimized by the Adam optimizer with the learning rate of $0.001$ and the decaying rate parameters $\beta_1 = 0.9$ and $\beta_2 = 0.999$. 
The training is continued either until the validation loss, measured using root MSE (RMSE), reaches 2.7\,dB, or up to 100 training epochs (156 stochastic gradient descent steps). The validation is perfomed after each epoch.


\textit{\textbf{Wireless Channel Parameters.}} Our wireless communication environment is described by the following parameters: $P^{(\mathrm{UL})} = 7.5$\,dBm, $P^{\mathrm{(DL)}} = 40$\,dBm, $W^{\mathrm{(DL)}} = 100$\,MHz and $W^{\mathrm{(UL)}} = 30$\,MHz, $r = 4$\,m, $\alpha = 5$, $\tau = 1$\,ms, and $\sigma^2 = -174$\,dBm/Hz.

\begin{table}[t]
  \caption{Privacy leakage  and success probability.}
  \vspace{- 1em}
  \label{table:leak_prob}
    \centering
    \resizebox{\columnwidth}{!}{\begin{tabular}{rcccc}\toprule
      & \multicolumn{4}{c}{Pooling dimension $w_{\mathrm{H}}\times w_{\mathrm{W}}$ (pixels)}\\
      &\hspace{10pt} $1 \times 1$ \hspace{10pt}&\hspace{10pt} $4 \times 4$\hspace{10pt} & \hspace{10pt}$10 \times 10$ \hspace{10pt}&  $40 \times 40$ (\textbf{1-pixel})\\\midrule
     Privacy leakage & 0.353 & 0.343 & 0.333 &  \textbf{0.296}\\
     Success Probability & 0.00 & 0.0270 & 0.999& \textbf{1.00} \\\bottomrule
    \end{tabular}}
    \vspace{-1.5em}
\end{table}

\textbf{Results and Discussions}
Fig.~\ref{fig:results}a illustrates the learning curves of our multimodal SL based on depth-images and RF signals (\textsf{Img+RF}), compared to two baselines based solely on images (\textsf{Img}) and RF signals (\textsf{RF}), respectively. The result shows that \textsf{RF} yields faster convergence thanks to its low sample complexity induced by the low-dimensional samples at the cost of lower accuracy. 
Except for \textsf{RF}, 1-pixel \textsf{Img+RF} (i.e., $40\times40$ pooling) achieves both the fastest convergence and highest accuracy, even outperforming the \textsf{Img+RF} with less CNN output compression (i.e., $4\times4$ pooling). 
This comes from a more compressed feed-forward payload that saves time for the transmissions, thereby enabling more computing iterations.

Fig.~\ref{fig:results}b visualizes the predicted received power and the ground-truth. It shows that \textsf{RF} performs well in LOS conditions, whereas \textsf{Img} is good at predicting the transitions between LoS and non-LoS conditions.
\textsf{Img+RF} inherits the benefits from both baseline schemes, achieving the closest prediction to the ground-truth.

Finally, Table~\ref{table:leak_prob} describes the privacy leakage and feed-forward signal decoding success probability. 
The privacy leakage is quantified with the inverse of the similarity between each raw image sample $x_k$ and its feature map $\phi(x_k)$ at the CNN output layer measured by multidimensional scaling algorithm \cite{hout2016using}.
For different pooling dimensions, the 1-pixel \textsf{Img+RF} achieves minimum privacy leakage and maximum success probability, owing to maximally compressing the CNN output images, as visualized in Fig.~\ref{fig:images}.

\section{Concluding Remarks}
We proposed a communication-efficient and privacy-preserving SL framework that exploits preceding RF signals and depth-images for mmWave received power prediction. 
\\

\textbf{Acknowledgement.}
The author would like to thank Mr.\ Kota Nakashima for providing the data set.
This work was supported in part by JSPS KAKENHI Grant Numbers JP17H03266, JP18H01442, and KDDI Foundation.
This work was also supported in part by the Academy of Finland under Grant 294128, in part by the 6Genesis Flagship under Grant 318927, in part by the KvantumInstitute Strategic Project (SAFARI), in part by the
Academy of Finland through the MISSION Project under Grant 319759, and in part by the Artificial Intelligence for Mobile Wireless Systems (AIMS) project at the University of Oulu.

\bibliographystyle{ACM-Reference-Format}
\bibliography{conext19}

\end{document}